\documentclass[a4paper,12pt]{article}
\usepackage[pctex32]{graphics}
\usepackage{amssymb,amsmath}

\begin{document}
\topmargin 0pt
\oddsidemargin 0mm
\newcommand{\be}{\begin{equation}}
\newcommand{\ee}{\end{equation}}
\newcommand{\ba}{\begin{eqnarray}}
\newcommand{\ea}{\end{eqnarray}}
\newcommand{\fr}{\frac}
\renewcommand{\thefootnote}{\fnsymbol{footnote}}

\begin{titlepage}

\vspace{5mm}
\begin{center}
{\Large \bf Nonexistence of quasinormal modes \\ in the extremal BTZ
black hole}

\vskip .6cm
 \centerline{\large
 Yun Soo Myung$^{1,a}$, Yong-Wan Kim $^{1,b}$,
and Young-Jai Park$^{2,c}$}

\vskip .6cm

{$^{1}$Institute of Basic Science and School of Computer Aided
Science, \\Inje University, Gimhae 621-749, Korea \\}

{$^{2}$Department of Physics and Department of Service Systems Management and Engineering, \\
Sogang University, Seoul 121-742, Korea}

\end{center}

\begin{center}

\underline{Abstract}
\end{center}
We show that quasinormal modes cannot exist in the extremal BTZ
black hole. For this purpose, we consider propagations of a
minimally coupled scalar and a single massive graviton obtained from
the cosmological topologically massive gravity on the extremal BTZ
black hole. The would-be quasinormal modes for  a scalar and
graviton could not exist because it is impossible to make an ingoing
flux into the extremal (degenerate) horizon. This is consistent with
the argument that there is no propagating dynamics in the self-dual
orbifold of AdS$_3$ which is just the near-horizon limit of the
extremal BTZ black hole.

 \vskip .6cm

\noindent PACS numbers: 04.70.Bw, 04.60.Kz, 04.30.Nk \\
\noindent Keywords: Quasinormal modes; extremal BTZ black hole;
topologically massive gravity \vskip 0.8cm

\vspace{15pt} \baselineskip=18pt
\noindent $^a$ysmyung@inje.ac.kr \\
\noindent $^b$ywkim65@gmail.com\\
\noindent $^c$yjpark@sogang.ac.kr

\thispagestyle{empty}
\end{titlepage}

\newpage
\section{Introduction}

It is well known that Einstein gravity in three dimensions has no
propagating degrees of freedom.   Massive generalizations of
three-dimensional gravity  allow propagating degrees of freedom.
Topologically massive gravity (TMG) is the famous gravity theory
obtained by including  a gravitational Chern--Simons term with
coupling $\mu$~\cite{DJT, DJT2}. The model was extended by the
addition of a cosmological constant $\Lambda=-1/\ell^2$  to the
topologically massive gravity (CTMG)~\cite{Deser82}. Since the
gravitational Chern--Simons term is odd under parity, the theory
shows  a single massive propagating degree of freedom of a given
helicity, whereas the other helicity mode remains massless. The
single massive field is realized as a massive scalar
$\varphi=z^{3/2}h_{zz}$ when using the Poincare coordinates
$x^{\pm}$ and $z$  covering the AdS$_3$ spacetimes~\cite{CDWW}.
However, it was claimed that the massive graviton having
negative-energy disappears at the critical point of
$\mu\ell=1$~\cite{LSS}. This cosmological topological massive
gravity at the critical point (CCTMG) may be described by the
logarithmic conformal field theory (LCFT)~\cite{GJ,Myung} even for
the zero central charge $c_L=0$. Bergshoeff, Hohm, and Townsend
recently proposed another massive generalization of Einstein gravity
by adding a specific quadratic curvature term to the
Einstein-Hilbert action~\cite{bht,bht2}. This term was designed to
reproduce  the ghost-free Fierz-Pauli action for a massive
propagating graviton in the linearized approximation. This gravity
theory became known as new massive gravity (NMG). Unlike the TMG,
the NMG preserves parity. As a result, the gravitons acquire the
same mass for both helicity states, indicating two massive
propagating degrees of freedom.

On the other hand, the BTZ black hole~\cite{BTZ1,BTZ2} as solution
to Einstein gravity with $\Lambda$ is also a black hole solution to
CTMG. Its quasinormal modes (QNMs) was calculated
in~\cite{Chan,Cardoso} and the CFT approach appeared in~\cite{BSS1}.
However, this does not necessarily imply that there is no difference
in the dynamics of perturbations. It is obvious that the
perturbation discriminates between Einstein gravity and CTMG.
Moreover, it is worth noting that the QNMs for the tensor
perturbation were shown to be the same as those derived by a massive
scalar when using the operator method~\cite{SS}. The asymptotic
properties of CTMG were studied in~\cite{Hen1,Hen2}.  Recently, it
was shown that the non-rotating BTZ black hole is stable for all
values of coupling $\mu$ against the metric perturbations in CTMG by
showing the presence of  left-and right-moving normal
modes~\cite{Birm}. Very recently, we have checked the stability of
the non-rotating BTZ black hole in the NMG by computing quasinormal
modes~\cite{mkp2}. This indicates that a minimally coupled massive
scalar plays a role of the barometer in finding quasinormal modes of
the tensor perturbation in the BTZ black hole background.

It is well known that on the contrary to the non-rotating BTZ black
hole, the QNMs of the extremal BTZ black hole do not exist for
scalar and fermionic perturbations~\cite{CLS}. Similarly, the QNMs
of the massless BTZ black hole did not exist for scalar perturbation
~\cite{ML} and fermionic perturbation~\cite{RWY}. It was suggested
that the absence of QNMs of extremal BTZ black hole is closely
related to the non-dynamical propagations in  the near-horizon limit
(a self-dual orbifold of AdS$_3$) of the extremal BTZ black
hole~\cite{BBSS,BPR,BSS}.

However, there were two works which report that the QNMs of the
extremal BTZ black hole are found for the scalar
perturbation~\cite{CZ} and tensor perturbation~\cite{AAM} in CTMG.
Let us call these the would-be  QNMs.

In this work, we confirm that  the would-be QNMs of the extremal BTZ
black hole do not exist for scalar and tensor perturbations by
showing that there is no ingoing flux onto the extremal horizon. For
this purpose, we introduce the Gaussian Normal coordinates
($u,v,\rho$) to simplify the extremal BTZ geometry.

\section{Scalar Propagation}

It is known that the extremal BTZ black hole as a solution of
three-dimensional Einstein gravity is described by the Schwarzschild
coordinates $(t,r,\tilde{\phi})$ as \be ds^2_{\rm
EBTZ}=g_{\mu\nu}dx^\mu
dx^\nu=-\Big(\frac{r^2}{\ell^2}-2\frac{r_{ex}^2}{\ell^2}\Big)dt^2
+\frac{r^2\ell^2}{(r^2-r_{ex}^2)^2}dr^2-2\frac{r_{ex}^2}{\ell}dt
d\tilde{\phi} +r^2d\tilde{\phi}^2, \ee whose degenerate horizon
$r=r_{ex}$ is determined by $g^{rr}=0$.  For our purpose, we could
express it in terms of the Gaussian Normal coordinates ($u,v,\rho$)
 \be\label{ebtz-metric}
 ds^2_{\rm EBTZ}=\bar{g}_{\mu\nu}dx^{\mu}dx^{\nu}=r_{ex}^2du^2-\ell^2e^{2{\rho}}du dv+\ell^2d{\rho}^2,
 \ee
where $u=t/\ell+\tilde{\phi}$, $v=t/\ell-\tilde{\phi}$, and
$\ell^2e^{2\rho}=r^2-r^2_{ex}$. For simplicity, we choose $\ell=1$.
In this coordinate system, the location of  horizon ($r=r_{ex}$)
corresponds to $\rho=-\infty$, while the infinity of $r=\infty$
corresponds to $\rho=\infty$. Note that the extremal BTZ spacetime
has the asymptotically AdS$_3$ spacetime.

Now, we are ready to study linear perturbations first for a scalar
field in this section, and then for a tensor field in the next
section by considering  the metric perturbation $h_{\mu\nu}$
 \be
 g_{\mu\nu}=\bar{g}_{\mu\nu}+h_{\mu\nu}.
 \ee
The reason why one  considers the scalar field first is clearly
understood when realizing that the QNMs of a minimally coupled
scalar usually provides a prototype of tensor QNMs to any black
holes~\cite{Berti,KZ}. Then, let us consider a massive scalar field
with a mass $\mu$, whose equation of motion is described  by
 \be \label{eom00}
 (\bar{\nabla}^2-\mu^2)\Phi=0,
 \ee
  where the overbar ( $\bar{}$ ) means the extremal BTZ background.
Considering the background symmetry, one has the ansatz
 \be
 \Phi(t,r,\phi)=e^{-i\omega t- ik\tilde{\phi}}\varphi(r),
 \ee
which leads to with $h/\bar{h}=(\omega \pm k)/2$
 \be
 \Phi(u,v,\rho)= e^{-ihu-i\bar{h}v}\varphi(\rho).
 \ee
Then, the equation of motion becomes
 \be \label{eom01}
 \varphi''(\rho)+2\varphi'(\rho)+4e^{-2\rho}(h\bar{h}+\bar{h}^2r^2_{ex}e^{-2\rho})\varphi(\rho)-\mu^2\varphi(\rho)=0,
 \ee
 where the prime denotes the
differentiation with respect to $\rho$. If one redefines
$\rho$-coordinate with a new  $z$ as
 \be
 z=2\bar{h}r_{ex}e^{-2\rho},
 \ee
Eq. (\ref{eom01}) leads to the Schr\"odinger-type equation
 \be\label{2ndeomphi}
 \frac{d^2\varphi}{dz^2}+\Big[E-V_\varphi(z)\Big]\varphi=0,
 \ee
where the energy $E$ and potential $V_\varphi$ are given by
 \be
 E=\frac{1}{4},~~~V_\varphi(z)=\frac{\mu^2}{4z^2}-\frac{\lambda}{z}
 \ee
with $\lambda=\frac{h}{2r_{ex}}$. Note that the potential depicted
in Fig.~1 shows that it grows to infinity as $z \to 0~(r\to
\infty$), while it approaches slowly zero as $z \to \infty~ (r \to
r_{ex})$.

\begin{figure}[t!]
   \centering
   \includegraphics{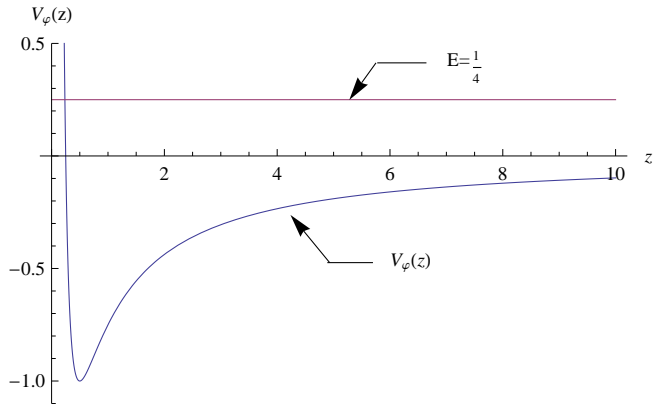}
\caption{Potential $V_\varphi(z)$ graph as function of $z$ for
$\lambda=\mu=1$.}
\end{figure}
In order to find the QNMs of a scalar field propagating the extremal
BTZ spacetime, one requires the boundary condition: the normalizable
mode at $z=0$ ($r=\infty$) and ingoing mode at $z=\infty$
($r=r_{ex}$). Observing the potential and energy leads to the
elementary quantum mechanics to determine the wave function at two
boundaries.  Near $z\sim 0$ ($r\rightarrow\infty$), its normalizable
solution for $\mu>0$ is
 \be
 \varphi_0\sim
 z^{\tilde{s}_+},~~\tilde{s}_+=\frac{1+\sqrt{\mu^2+1}}{2}
 \ee
while for $z\rightarrow\infty$ ($r\rightarrow r_{ex}$), its incoming
solution is described by
 \be
 \varphi_\infty \sim e^{\frac{i}{2}z}.
 \ee
On the other hand, the intermediate solution between $z\sim 0$ and
$z\rightarrow\infty$ is described by taking
$\varphi=\varphi_0\varphi_\infty f_\Phi(z)$, where $f_\Phi(z)$
satisfies the differential equation
 \be \label{scalareq2}
 f_\Phi''(z)+\left(\frac{2\tilde{s}_++iz}{z}\right)f_\Phi'(z)
       +\left(\frac{i\tilde{s}_++\lambda}{z}\right)f_\Phi(z)=0.
 \ee
Redefining $\xi=-iz$, this differential equation  becomes
 \be \label{scalareq3}
 \xi\frac{d^2f_\Phi(\xi)}{d\xi^2}+(2\tilde{s}_+-\xi)
  \frac{d f_\Phi(\xi)}{d\xi}-\left(\tilde{s}_+-i\lambda\right)f_\Phi(\xi)=0
 \ee
whose normalizable solution is determined to be
 \be
 f_\Phi(\xi)\sim F[\tilde{s}_+-i\lambda,2\tilde{s}_+;\xi].
 \ee
Here $F[a,c;\xi]$ is the confluent hypergeometric
function~\cite{ASS}.

As a result, we have the full solution to Eq. (\ref{2ndeomphi})
written as
 \be
 \Phi(u,v,z) \sim e^{-ihu-i\bar{h}v} \varphi(z)
 \ee
 with
 \be
 \varphi(z)=z^{\tilde{s}_+} e^{\frac{i}{2}z}
            F[\tilde{s}_+-i\lambda,2\tilde{s}
            _+;\xi].
 \ee
We will derive  the would-be QNMs  in the end of the following
section because the solution $\Phi(u,v,z)$ has the nearly same form
as that of a propagating tensor mode $h_{vv}$.

\section{Tensor Propagation}

In this section, we study the linear perturbation for a tensor field
in the extremal BTZ background. The action of TMG is given by
 \be
 I_{TMG}=\frac{1}{\kappa^2}\left(I_{EH}+\frac{1}{\mu}I_{CS}\right),
 \ee
with the Einsten-Hilbert (EH) and the gravitational Chern-Simons
(CS) action
 \ba
 I_{EH}&=&\int d^3 x \sqrt{-g} \left(R+\frac{2}{\ell^2}\right),\nonumber\\
 I_{CS}&=& \frac{1}{2}\int d^3x\sqrt{-g}\epsilon^{\lambda\mu\nu}\Gamma^{\rho}_{\lambda\sigma}
  \left(\partial_{\mu}\Gamma^{\sigma}_{\rho\nu}+\frac{2}{3}\Gamma^{\sigma}_{\mu\tau}\Gamma^{\tau}_{\nu\rho}\right).
 \ea
Here, we denote $\kappa^2=16\pi G$, the cosmological constant,
$\Lambda=-1/\ell^2$, and a coupling constant, $\mu$. The equation of
motion of the TMG is obtained as
 \be
 G_{\mu\nu}+\frac{1}{\mu}C_{\mu\nu}=0,
 \ee
where the Einstein tensor $G_{\mu\nu}$ and the Cotton tensor
$C_{\mu\nu}$ are given by
 \ba
 G_{\mu\nu} &=& R_{\mu\nu}-\frac{1}{2}g_{\mu\nu}R+\Lambda g_{\mu\nu},\nonumber\\
 C_{\mu\nu} &=& \epsilon^{\alpha\beta}_\mu\nabla_\alpha
                \left(R_{\beta\nu}-\frac{1}{4}g_{\beta\nu}R\right),
 \ea
respectively. As was mentioned in the introduction, the TMG
provides  a single massive propagating mode.

The perturbed Einstein equation  of the TMG under the
transverse-traceless (TT) gauge ($\bar{\nabla}^\mu
h_{\mu\nu}=0,~h^\rho~_\rho=0$) leads to the third-order
equation~\cite{LSS}
 \be
 \left(\bar{\nabla}^2+\frac{2}{\ell^2}\right)
 \left(h_{\mu\nu}+\frac{1}{\mu}\epsilon_\mu^{~\alpha\beta}\bar{\nabla}_\alpha
 h_{\beta\nu}\right)=0,
 \ee
 where the overbar ( $\bar{}$ ) means the extremal BTZ background.
However, when considering the TT gauge and massive propagation only,
it is enough to consider the first-order equation which describes a
massive graviton
 \be \label{1steq}
  \mu h_{\mu\nu}+\epsilon_\mu^{~\alpha\beta}\bar{\nabla}_\alpha h_{\beta\nu}=0
 \ee
 because $\left(\bar{\nabla}^2+\frac{2}{\ell^2}\right)h_{\mu\nu}=0$
describes the massless graviton, being gauge-artefact in three
dimensions. Starting with  six components of a symmetric tensor as
 \ba
 h_{\mu\nu} = e^{-ihu-i\bar{h}v}
   \left(\begin{array}{ccc}
       F_{uu}(\rho) & F_{uv}(\rho) & F_{u\rho}(\rho) \\
       F_{uv}(\rho) & F_{vv}(\rho) & F_{v\rho}(\rho) \\
       F_{u\rho}(\rho) &  F_{v\rho}(\rho) & F_{\rho\rho}(\rho)
       \end{array}
  \right),
 \ea
we obtain  two first-order differential equations from (\ref{1steq})
 \ba\label{eom1}
 F'_{vv}&=&-(\mu-1)F_{vv}-i\bar{h}F_{\rho v}, \\
 \label{eom2}
 F'_{\rho v}&=&-(\mu+1)F_{\rho v}-i\bar{h}F_{\rho\rho},
 \ea
and four algebraic relations
 \ba
 (\mu+1)F_{\rho u}e^{2\rho}&=&2i(hF_{uv}-\bar{h}F_{uu}),\nonumber\\
 \mu F_{\rho\rho}e^{4\rho}&=&2i(hF_{\rho v}-\bar{h}F_{\rho u})e^{2\rho}+4F_{vv}r_{ex}^2,\nonumber\\
 (\mu-1)F_{\rho v}e^{2\rho}&=&2i(hF_{vv}-\bar{h}F_{uv}),\nonumber\\
 F_{\rho\rho}e^{4\rho}&=&4(r_{ex}^2F_{vv}+e^{2\rho}F_{uv}).
 \ea
Then, one may rewrite all other components in terms of $F_{vv}$ and
$F'_{vv}$ as
 \ba
 F_{v\rho}&=&\frac{i}{\bar{h}}\left[(\mu-1)F_{vv}+F'_{vv}\right],\\
 F_{uv}&=&-\frac{1}{2\bar{h}^2}\left[((\mu-1)^2e^{2\rho}-2h\bar{h})F_{vv}+(\mu-1)e^{2\rho}F'_{vv}\right],\\
 F_{\rho\rho}&=&-\frac{2}{\bar{h}^2}\left[((\mu-1)^2-2\bar{h}^2r_{ex}^2
                   e^{-4\rho}-2h\bar{h}e^{-2\rho})F_{vv}+(\mu-1)F'_{vv}\right],\\
 F_{u\rho}&=&-\frac{i}{\bar{h}^3}\left[(\mu(\mu-1)^2e^{2\rho}-2\bar{h}^2r_{ex}^2(\mu-1)e^{-2\rho}
                 +h\bar{h}(1-3\mu))F_{vv}\right. \nonumber\\
          && ~~~~~~ \left.+(\mu(\mu-1)e^{2\rho}-h\bar{h})F'_{vv}\right],\\
 F_{uu}&=&\frac{1}{2\bar{h}^4}\big[\left(\mu(\mu+1)(\mu-1)^2e^{4\rho}-4\mu^2
                  h\bar{h}e^{2\rho}+2(h^2-r_{ex}^2(\mu^2-1))\bar{h}^2\right)F_{vv}      \nonumber \\
       &&~~~~~~+\mu\left((\mu^2-1)e^{2\rho}-2h\bar{h}\right)e^{2\rho}F_{vv}'\big],
 \ea
which show that  a single propagating mode  is $F_{vv}$.

On the other hand, since the first-order equations are not suitable
to derive the QNMs,  we need the second-order equation for $F_{vv}$.
Differentiating Eq. (\ref{eom1}) with respect to $\rho$, and
replacing $F'_{vv}$, and  $F'_{\rho v}$ again, one arrives  at the
second-order differential equation~\cite{AAM}
 \be \label{graveq}
 F_{vv}''+2F_{vv}'+4(\bar{h}h+\bar{h}^2r_{ex}^2e^{-2\rho})e^{-2\rho}F_{vv}
 -(\mu-1)(\mu-3)F_{vv}=0,
 \ee
which is the same in Eq. (\ref{eom01}) except the last mass term.
Hence, we could represent the graviton equation (\ref{graveq})
effectively as the massive scalar equation
 \be
 \Big[\bar{\nabla}^2-(\mu-1)(\mu-3)\Big]F_{vv}=0,
 \ee
which explains clearly  why Eq. (\ref{eom00}) is considered as a
prototype of the tensor-perturbed equation.

Introducing
 \be
 z=2\bar{h}r_{ex}e^{-2\rho},
 \ee
this equation becomes the Schr\"odinger-like equation
 \be\label{2ndeomh}
 \frac{d^2F_{vv}}{dz^2}+\left[E-V_{h}(z)\right]F_{vv}=0
 \ee
with the energy $E$ and potential $V_{h}(z)$
 \be
 E=\frac{1}{4},~~~V_{h}(z)=\frac{m^2-\frac{1}{4}}{z^2}-\frac{\lambda}{z}.
 \ee
Here $m^2=(\mu/2-1)^2$ and  $\lambda=\frac{h}{2r_{ex}}$. Hereafter,
we choose $m=\frac{\mu}{2}-1$ without loss of generality. The
potential is depicted in Fig.~2. We wish to  point out that the
shape of potential $V_{h}$ is very similar to
 the scalar potential $V_\varphi$, which means that their asymptotic
 forms are the same but the difference is the depth of their potentials.
 Importantly, their energies are the same.
This implies that two fields $\Phi$ and $h_{vv}$ may provide the
nearly same QNMs if they exist.

\begin{figure}[t!]
   \centering
   \includegraphics{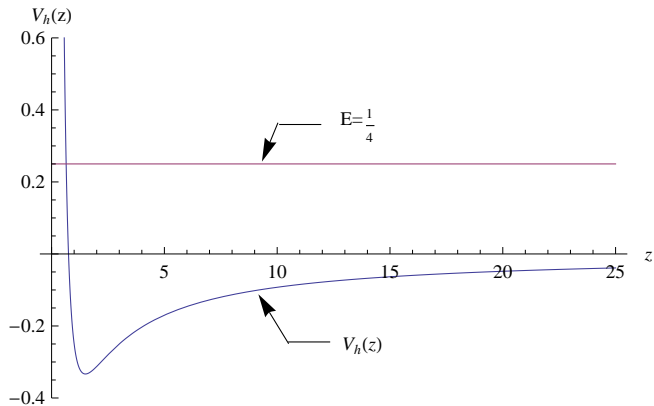}
\caption{Potential $V_h(z)$ graph as function of $z$ for $\mu=4$ and
$\lambda=1$.}
\end{figure}
In order to find the QNMs of a tensor mode $h_{vv}$, we require the
boundary condition.  Near $z\sim 0$ ($r\rightarrow\infty$), its
normalizable solution for $m>1/2$ is
 \be
 F^0_{vv}\sim z^{m+\frac{1}{2}}=z^{s_+},~~s_+=\frac{\mu-1}{2}.
 \ee
On the other hand, when $z\rightarrow\infty$ ($r\rightarrow
r_{ex}$), its ingoing solution is
 \be
 F^\infty_{vv}\sim e^{\frac{i}{2}z}.
 \ee
To obtain a full solution in the whole region, we take
$F_{vv}=F^0_{vv}F^\infty_{vv}f_h(z)$, and insert it into Eq.
(\ref{2ndeomh}). Then, the function $f_h(z)$ connecting between the
near horizon and the asymptotic infinity  satisfies
 \be
 f_h''(z)+\frac{2s_++iz}{z}f_h'(z)+\frac{is_++\lambda}{z}f_h(z)=0.
 \ee
Now, defining $\xi=-iz$, we have
 \be
 \xi \frac{d^2f_h(\xi)}{d\xi^2}+(2s_+-\xi)\frac{d f_h(\xi)}{d\xi}
   -\left(s_+-i\lambda\right)f_h(\xi)=0,
 \ee
 which leads to the same equation as in Eq. (\ref{scalareq3}) when replacing
 $s_+$ by $\tilde{s}_+$.
Hence, its normalizable solution is given by the confluent
hypergeometric function
 \be
 f_h(\xi)\sim F[s_+-i\lambda,2s_+,\xi].
 \ee
Finally, we  have the full solution as
 \be \label{normal-sol}
 h_{vv} \sim e^{-ihu-i\bar{h}v} F_{vv}(z)
 \ee
 with
 \be
F_{vv}(z)= z^{s_+} e^{\frac{i}{2}z}
 F[s_+-i\lambda,2s_+,\xi]
 \ee
which indicates   the normalizable solution near $z\sim 0$ and the
ingoing mode at $z\rightarrow\infty$.

Before making a further analysis,  we observe the  useful property
for the complex conjugate of the confluent hypergeometric function
as
 \be \label{cchg}
 F^*[a,c;\xi]=F[a^*,c;-\xi],
 \ee
where $a=s_+-i\lambda$ and  $c=2s_+$. Together with the Kummer's
transformation of $F[a,c;\xi]=e^\xi F[c-a,c;-\xi]$, Eq. (\ref{cchg})
implies that the product of $e^{\frac{i}{2}z}$ and $F[a,c;\xi]$ is
real as
 \be
 \left[e^{\frac{i}{2}z} F[a,c;\xi]\right]^*
 =e^{\frac{i}{2}z} F[a,c;\xi].
 \ee
Now, we are in a position to calculate the radial flux defined by
 \be
 {\cal F}_\phi=2\frac{2\pi}{i}\sqrt{-g}[\phi^*\partial_\rho\phi-\phi\partial_\rho\phi^*]
 \rightarrow 8\pi i\bar{h}r_{ex}[\phi^*\partial_z\phi-\phi\partial_z\phi^*]
 \ee
for a proper mode solution $\phi\in\{\varphi(z),F_{vv}(z)\}$.

Firstly, we calculate the flux near $z\sim 0$
($r\rightarrow\infty$). For $z\rightarrow 0$, it is clear that
$F[s_+-i\lambda,2s_+,\xi]\rightarrow 1$ and
$e^{\frac{i}{2}z}\rightarrow 1$.  Thus, the full normalizable
solution reduces to
 \be
 F_{vv} \to  F^0_{vv} \sim z^{s_+},
 \ee
which makes the flux zero ( ${\cal F}|_{z\rightarrow 0}=0$) because
$F^0_{vv}$ is real. This is consistent with the Dirichlet boundary
condition at infinity of AdS$_3$ spacetimes.

Secondly, we compute the ingoing flux near the extremal horizon at
$z\rightarrow\infty$ ( $r\rightarrow r_{ex}$). Since
$e^{\frac{i}{2}z} F[s_+-i\lambda,2s_+,\xi]$ is real, the flux of
${\cal F}|_{z\rightarrow\infty}$ is simply  zero. On the other hand,
if one uses $F^\infty_{vv}\sim e^{\frac{i}{2}z}$ to compute the
ingoing flux, one has the non-zero flux of  ${\cal F}|_{z\rightarrow
\infty}=-8\pi \hbar r_{ex}$. This contradiction arises because we do
not develop an explicit form of wave function in the near-horizon
geometry of the extremal BTZ black hole.  In order to obtain  the
desired wave function in the near-horizon region,  we use  the
expansion formula of the confluent hypergeometric function  near the
extremal  horizon of $z\to \infty$. For large $|\xi|$,  one
has~\cite{ASS}
 \be
 F[a,c;\xi]=\frac{\Gamma(c)}{\Gamma(c-a)}e^{\pm i\pi a}\xi^{-a}
            +\frac{\Gamma(c)}{\Gamma(a)}e^\xi \xi^{a-c},
 \ee
where the upper sign is for $-\pi/2< {\rm arg}(\xi)<3\pi/2 $ and the
lower sign is for $-3\pi/2< {\rm arg}(\xi) \le -\pi/2$. Since we use
$\xi=-iz$ here, we take the lower sign in the asymptotic expansion
which becomes explicitly
 \be
 F[a,c;\xi]=\frac{\Gamma(c)}{\Gamma(c-a)}|\xi|^{-a}e^{-i\pi a/2}
            +\frac{\Gamma(c)}{\Gamma(a)}e^\xi |\xi|^{a-c}e^{i\pi(c-a)/2}.
 \ee
As a result, one finds that the explicit form of wave function
 \ba
 e^{\frac{i}{2}z} F[a,c;\xi]|_{z\to \infty} &\sim&
            \left[\frac{\Gamma(2s_+)}{\Gamma(s_++i\lambda)}
            e^{-\frac{\pi}{2}\lambda}e^{i(\frac{z}{2}+\lambda\ln z-\frac{\pi}{2}s_+)}
            \right. \nonumber\\
         &&~~~~~~~~~~~~ \left. +\frac{\Gamma(2s_+)}{\Gamma(s_+-i\lambda)}
            e^{-\frac{\pi}{2}\lambda}e^{-i(\frac{z}{2}+\lambda\ln z-\frac{\pi}{2}s_+)}
            \right]\nonumber  \\
            &\equiv& F^{in}_{vv}+ F^{out}_{vv}
 \ea
in the near-horizon region of the extremal BTZ black hole. We note
that the first term is the ingoing mode ($\to$), while the second
term is the outgoing mode ($\leftarrow$) near $z=\infty$
($r=r_{ex}$). Importantly, we confirm  that $e^{\frac{i}{2}z}
F[a,c;\xi]|_{z\to \infty}$ is real because of
$[F^{in}_{vv}]^*=F^{out}_{vv}$. In order to obtain the QNMs, the
wave function should be purely ingoing mode near the extremal
horizon.  This may be done by requiring $s_+-i\lambda=-n$,
($n=1,2,3,\cdot\cdot\cdot$), which amounts to taking the ingoing
flux without outgoing flux.  In this case,  we may obtain  the
would-be QNMs of a tensor mode $h_{vv}$ from the condition of
$s_+-i\lambda=-n$ with $\lambda=h/2r_{ex}$ and $s_+=(\mu-1)/2$
 \be
 \omega_h=-k-i4r_{ex} \Big(n+s_+\Big),
 \ee
which was exactly  the same QNMs found in~\cite{AAM}. However, since
the corresponding ingoing-radial flux\footnote{More precisely, we
have a factor of $\left(1+\frac{2\lambda}{z}\right)$ in the front of
this expression. However, in the limit of $z\to \infty$, this
reduces to 1.} is zero  when requiring the condition of the
no-outgoing mode ($s_+-i\lambda=-n$) as
 \ba
 {\cal F}_h^{\rm in}(z\rightarrow\infty) &=&
     8\pi i \bar{h}r_{ex}[F^{in*}_{vv}\partial_z F^{in}_{vv}-F^{in}_{vv}\partial_z F^{in*}_{vv}]\nonumber\\
     &=& -8\pi  \bar{h}r_{ex} e^{-\pi \lambda}
      \left[\frac{\Gamma(2s_+)}{\Gamma(s_+-i\lambda)}
            \frac{\Gamma(2s_+)}{\Gamma(s_++i\lambda)}\right]=0,
 \ea
it concludes that there exist no QNMs for the tensor perturbation in
the extremal BTZ background. Here, $\Gamma(-n)=\infty$, and  the
minus sign means that it is the ingoing flux near
$z\rightarrow\infty$.

The same thing happens to the massive scalar mode by replacing $s_+$
by $\tilde{s}_+$. We may take the would-be QNMs of a scalar mode
$\Phi$
 \be
 \omega_\Phi=-k-i4\pi T_L\Big(n+\tilde{s}_+\Big),
 \ee
which is exactly the same QNMs found in~\cite{CZ} with the
left-temperature $T_L=r_{ex}/\pi$ and
$\tilde{s}_+=(1+\sqrt{\mu^2+1})/2$. However, its ingoing flux is
zero when requiring $\tilde{s}_+-i\lambda=-n$ \ba
 {\cal F}_\Phi^{\rm in}(z\rightarrow\infty) &\sim&
     8\pi i \bar{h}r_{ex}[\varphi^{in*}\partial_z \varphi^{in}-\varphi^{in}\partial_z \varphi^{in*}]\nonumber\\
     &=&- 8\pi  \bar{h}r_{ex} e^{-\pi \lambda}
      \left[\frac{\Gamma(2\tilde{s}_+)}{\Gamma(\tilde{s}_+-i\lambda)}
            \frac{\Gamma(2\tilde{s}_+)}{\Gamma(\tilde{s}_++i\lambda)}\right]=0,
 \ea
which implies that there is no QNMs of scalar perturbation, too.
This consists with the previous works~\cite{CLS,ML}  for a massive
scalar propagation. Also, the absence of QNMs is consistent with the
argument that there is no propagating dynamics in the self-dual
orbifold of AdS$_3$, which is just the near-horizon limit of the
extremal BTZ black hole~\cite{BBSS,BPR,BSS}.

\section{Discussions}
In this work we have shown that the would-be quasinormal modes of a
massive scalar and a single massive graviton do not exist in the
extremal BTZ black hole. This shows the contradiction  to the
previous results on the scalar propagation~\cite{CZ} and a graviton
propagation in the TMG~\cite{AAM}.

It was believed that the extremal BTZ black hole is a
non-dissipative system because its thermodynamic quantities are
characterized  by the zero temperature and heat capacity
$T_H=C_J=0$, but the non-zero entropy $S_{BH}=\frac{\pi r_{ex}}{2G}$
[$S_{BH}=\frac{\pi r_{ex}}{2G}(1+\frac{1}{\mu})$ for the TMG].
Actually, two propagating equations provide  the nearly same
Schr\"odinger-type equations (\ref{2ndeomphi}) and (\ref{2ndeomh})
with the same energy $E=1/4$. If the Schr\"odinger operator ${\cal
L}=-d^2/dz^2+V(z)$ is self-adjoint (${\cal L}^\dagger={\cal L}$),
its eigenvalue is real upon imposing the AdS$_3$-boundary condition.
In this case, there is no information loss via either evaporation or
absorption process and thus, the unitarity is preserved. This is
consistent with the picture that the extremal BTZ black hole is a
final remnant, which never evaporates and absorbs any radiations. If
the quasinormal modes (complex $\omega$) are found from the black
hole in the AdS$_3$ spacetime, the black hole is regarded as a
dissipative system. Therefore, the ingoing flux is not zero at the
horizon and the flux is zero  at the infinity. However, the extremal
black hole including the massless BTZ black hole belongs to the
non-dissipate system, contrary to the dissipative system of the
non-extremal BTZ black hole including the non-rotating BTZ black
hole. Hence it is reasonable to consider that quasinormal modes
could not obtained from the extremal BTZ black hole~\cite{LO}. From
(\ref{normal-sol}), the normal mode solution of $h_{vv}\sim
e^{-i\omega t-ik \tilde{\phi}} F_{vv}(z)$ with real $\omega$ is
allowed only, showing that the extremal BTZ black hole is stable
against the external perturbations.

Initially, the BTZ black hole could be holographically described by
a dual CFT with both left- and right-moving
temperatures~\cite{BSS1}. Since the extremal BTZ black hole has the
zero Hawking temperature and zero right-temperature, it was believed
that one sector of the  CFT is frozen, while the other sector
survives with $T_L=r_{ex}/\pi$. In this case, some people may
consider that the would-be QNMs of extremal BTZ black hole
correspond to the operators perturbing the thermal equilibrium in
the dual chiral CFT. However, it was suggested  that there is no
propagating dynamics in the AdS$_2$ base of the self-dual orbifold
of AdS$_3$ (AdS$_2\times S^1$) which is just the near-horizon limit
of the extremal BTZ black hole~\cite{BBSS}.  The near-horizon limit
of the extremal black hole is truly  dual to the discrete light-cone
quantization (DLCQ) of a non-chiral (ordinary)  CFT. The kinematics
of DLCQ implies  that in a consistent quantum theory of gravity
around the extremal BTZ black hole, there is no dynamics in the
AdS$_2$. In other words, the description of extremal BTZ black hole
in terms of an AdS$_2$ throat requires asymptotic boundary
conditions eliminating AdS$_2$ excitations. How would the TMG around
the extremal BTZ black hole require the absence of a massive
graviton $h_{vv}$ including a massive scalar $\Phi$ in the AdS$_2$
base of AdS$_2\times S^1$? This  may be because any fluctuations
would cause the space to fragment, leading to the appearance of
multiple boundaries to the spacetimes~\cite{MMS}.

Finally, if one finds the retarded correlation function in the DLCQ
theory, according to the AdS/CFT dictionary one can confirm the
absence of QNMs of the extremal BTZ black hole because the QNMs can
read off from the location of the poles of the retarded correlation
function of the corresponding perturbations in the dual
CFT~\cite{BSS}. For the massless BTZ black hole, it was confirmed
from the CFT side~\cite{RWY} that there is no QNMs of a massive
scalar propagation~\cite{ML}.

\section*{Acknowledgement}
Two of us (Y. S. Myung and Y.-W. Kim) were
 supported by the National
Research Foundation of Korea (NRF) grant funded by the Korea
government (MEST) (No.2011-0027293). Y.-J. Park was supported by the
Korea Science and Engineering Foundation (KOSEF) grant funded by the
Korea government (MEST) through WCU Program (No. R31-20002).

\end{document}